\documentclass[preprint,12pt]{elsarticle}

\usepackage{amssymb}

\usepackage{graphicx}
\usepackage[T2A]{fontenc}
\usepackage{amsmath,amssymb,epsfig}
\usepackage{comment}
\usepackage{color,subfigure}

\def\a {\alpha}
\def\ve {\varepsilon}

\def\b {\beta}
\def\g {\gamma} \def \G {\Gamma}
\def\d {\delta} 
\def\l {\lambda} \def\L {\Lambda}
\def\m {\mu}
\def\n {\nu}
\def\p {\pi}
\def\r {\rho}
\def\s {\sigma}

\def\dd{{\rm d}}

\def\({\left(}
\def\){\right)}
\def\[{\left[}
\def\]{\right]}
\def\nn{\nonumber}

\def\bf{\textbf}

\def\beq {\begin{equation}}
\def\eq {\end{equation}}
\def\bea{\begin{eqnarray}}
\def\ea{\end{eqnarray}}
\def\eea{\end{eqnarray}}

\def\({\left(}
\def\){\right)}
\def\[{\left[}
\def\]{\right]}

\newcommand{\notd}[1] { \setbox0=\hbox{$#1$}
\dimen0=\wd0   \setbox1=\hbox{/} \dimen1=\wd1  \ifdim\dimen0>\dimen1
 \rlap{\hbox to \dimen0{\hfil/\hfil}}  #1 \else \rlap{\hbox to \dimen1{\hfil$#1$\hfil}}  /  \fi  }

\journal{Physics Letters B}

\begin{document}

\begin{frontmatter}



\title{Single meson contributions to the muon's anomalous magnetic moment}


\author[lab1,lab2,lab3]{Vladyslav Pauk}
\author[lab1,lab2]{Marc Vanderhaeghen}

\address[lab1]{PRISMA Cluster of Excellence, Johannes Gutenberg-Universit\"at,  Mainz, Germany}
\address[lab2]{Institut f\"ur Kernphysik, Johannes Gutenberg-Universit\"at,  Mainz, Germany}
\address[lab3]{Department of Physics, Taras Shevchenko National University of Kyiv, Ukraine}

\begin{abstract}
We develop the formalism to provide an improved estimate for the hadronic light-by-light correction 
to the muon's anomalous magnetic moment $a_\mu$, by considering single meson contributions beyond the leading pseudo-scalar mesons.  We incorporate available experimental input as well as constraints from 
light-by-light scattering sum rules to estimate the effects of axial-vector, scalar, and tensor mesons. We give numerical evaluations for the hadronic light-by-light contribution of these states to $a_\mu$. The presented formalism allows to further improve on these estimates, once new data for such meson states will become available. 
\end{abstract}


\end{frontmatter}













\section{Introduction}

\begin{figure}[h]
\centering
  \includegraphics[width=4cm]{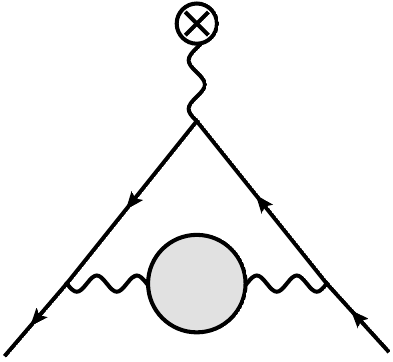}
  \hspace{2cm}
  \includegraphics[width=5cm]{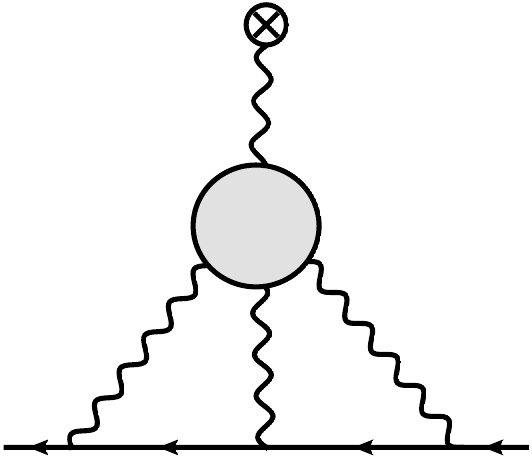}
  \caption{The hadronic contributions to the anomalous magnetic moment of the muon. Left panel: hadronic vacuum polarization (HVP). 
  Right panel: hadronic light-by-light contribution (HLbL). The grey blobs denote hadronic intermediate states.}
  \label{fig:hadronic}
\end{figure}

The anomalous magnetic moment of the muon $a_\mu = (g - 2)/2$ 
is one of the most precisely measured quantities in particle physics. 
It has being playing a vital role in testing the framework of quantum field theory since its development 
more than half a century ago, as well as in searching for new physics beyond the Standard Model (SM) of particles and interactions, for a comprehensive review, see~\cite{Jegerlehner:2009ry} and references therein.  
On the experimental side, the present world average for $a_\mu$ 
is~\cite{Bennett:2004pv, Bennett:2006fi}:
\begin{equation}
a_{\mu}(\mathrm{exp}) = ( 116 \, 592 \, 089 \pm 63 ) \times 10^{-11}, 
\end{equation} 
which corresponds to a relative precision of 0.54 parts per million. 
From the theoretical point of view, in the SM $a_\mu$ is defined by electromagnetic (QED), 
electroweak, and hadronic contributions. 
The dominant QED contribution, which at present has been calculated including all terms up to fifth-order in the fine structure constant~\cite{Aoyama:2012wk, Aoyama:2012fc}, is known to an impressive theoretical precision of  $\delta a_\mu (\mathrm{QED}) = 8 \times 10^{-13}$.  
The much smaller electroweak contribution, which has been calculated up to 2-loop order~\cite{Fujikawa:1972fe,Czarnecki:1995sz,Knecht:2002hr,Czarnecki:2002nt}, is also known with good accuracy 
$\delta a_\mu (\mathrm{weak}) = 2 \times 10^{-11}$, 
which is more than a factor of 30 smaller than the 
present experimental precision. 
Within the standard model, the largest source of uncertainty is given 
by the hadronic contribution, which contains two parts, the hadronic 
vacuum polarization (HVP) together with the 
hadronic light-by-light scattering (HLbL), see Fig.~\ref{fig:hadronic}. 
The HVP has been estimated based on data for $e^+ e^- \to \mathrm{hadrons}$, $e^+ e^- \to \gamma + \mathrm{hadrons}$, as well as $\tau$ decays, by  
several groups~\cite{Davier:2002dy, Benayoun:2007cu, Benayoun:2009im, Davier:2009zi, Davier:2009ag, Davier:2010nc, Jegerlehner:2011ti, Hagiwara:2011af, Benayoun:2012wc, Jegerlehner:2013sja}. 
A recent evaluation of the leading order HVP 
has found~\cite{Jegerlehner:2013sja}~:
\begin{equation}
 a_\mu (\mathrm{l.o. \, HVP}) = ( 6886.0 \pm 42.4) \times 10^{-11}. 
 \label{eq:hvp}
 \end{equation}
 The next-to-leading order HVP has been estimated as 
 ~\cite{Hagiwara:2011af}:
\begin{equation}
 a_\mu (\mathrm{n.l.o. \, HVP}) = ( -98.4 \pm 0.7) \times 10^{-11}. 
 \end{equation}
The HLbL, although much smaller in size than the HVP,  has a similar theoretical uncertainty. It has been estimated by different groups as~:
\begin{align}
 a_\mu (\mathrm{HLbL}) &= ( 116 \pm 39) \times 10^{-11} \quad \quad 
\mathrm{Ref.}~\mbox{\cite{Jegerlehner:2009ry}}, 
\label{eq:hlbl} \\
 a_\mu (\mathrm{HLbL}) &= ( 105 \pm 26) \times 10^{-11} \quad \quad 
 \mathrm{Ref.}~\mbox{\cite{Prades:2009tw}}. 
 \end{align}
When comparing theory with experiment for $a_\mu$, 
the difference has recently been  
evaluated as~\cite{Jegerlehner:2013sja}
\begin{equation}
 a_\mu (\mathrm{exp}) - a_\mu (\mathrm{theory}) = ( 312.5 \pm 57.6 \, \mathrm{(theory)} \pm 63 \, \mathrm{(exp)}) \times 10^{-11}, 
 \end{equation}
which corresponds with a $3.7 \sigma$ discrepancy. 
The different analyses for the l.o. HVP and HLbL contributions, give results 
which all agree within $1 \sigma$~\cite{Blum:2013xva}. 

In order to conclude whether this discrepancy is a sign of new physics 
beyond the standard model, new experiments are 
planned in the near future both at 
Fermilab~\cite{LeeRoberts:2011zz} as well as at 
J-PARC~\cite{Iinuma:2011zz} to further improve on the precision. 
The Fermilab experiment aims to reduce the experimental uncertainty 
by a factor 4 to $\delta a_\mu \approx 16 \times 10^{-11}$. 
Such improvement also calls to improve on the theoretical accuracy by 
at least a factor of 2 in order to obtain a definitive test for the presently observed 
discrepancy.  As the theoretical uncertainty is totally dominated by 
the knowledge of the HVP, Eq.~(\ref{eq:hvp}), 
and the HLbL, Eq.~(\ref{eq:hlbl}), the main effort on the theoretical side 
will be to improve on both estimates. For the HVP, new data from 
ongoing experiments at Novosibirsk and BES-III 
will provide valuable experimental input to further 
constrain this contribution. It was estimated in Ref.~\cite{Blum:2013xva} that 
such data will allow to reduce the uncertainty in the HVP to 
$\delta a_\mu (\mathrm{l.o. \, HVP}) = 26 \times 10^{-11}$.
For the HLbL scattering, new data are expected from KLOE-2 for the 
$\gamma^\ast \gamma \to \pi^0$ transition form factor  at very low photon virtualities, 
and from BES-III for the reactions $\gamma^\ast \gamma \to X$, where 
$X = \pi^0, \eta, \eta^\prime, 2 \pi$. 
Such data do require a theoretical analysis in order to further constrain 
the HLbL evaluation. 

The aim of the present work is to provide an improved estimate for the HLbL contribution, by considering single meson contributions beyond the leading pseudo-scalar mesons ($\pi^0, \eta, \eta^\prime$), 
which have been evaluated in the pioneering work of Ref.~\cite{Knecht:2001qf}.
We will incorporate available experimental input as well as constraints from 
light-by-light scattering sum rules~\cite{Pascalutsa:2010sj, Pascalutsa:2012pr}  
to estimate the effects of axial-vector, scalar, and tensor mesons to the HLbL contribution. 
The framework which will be presented will also allow to further improve on the estimate, once new data, in particular from BES-III, for such meson states will become available. 

\section{Meson pole contributions to the hadronic light-by-light scattering}

The HLbL contribution to the muon's anomalous magnetic moment is a $\mathcal{O}(\a^3)$ correction to  the Pauli form factor due to the second diagram in Fig.~\ref{fig:hadronic}. It may be isolated from 
the vertex matrix element 
\beq
\left<\m^-(p')\left|j_{em}^\m(0)\right|\m^-(p)\right>=(-ie)\bar{u}(p')\G^\m(p',p)u(p),
\eq 
where $p$ ($p^\prime$) denote the initial (final) muon momenta, 
when considering the classical limit ($p'-p \equiv k\to0$) by using a projection operator technique \cite{BS}. 
This amounts to the two-loop integral representation
\beq
\begin{split}
a_\m^{LbL}=&\lim\limits_{k\to0}ie^6\int\frac{\dd^4 q_1}{(2\p)^4}\int\frac{\dd^4 q_2}{(2\p)^4}\frac1{q_1^2q_2^2(k-q_1-q_2)^2}\frac1{(p+q_1)^2-m^2}\frac1{(p'-q_2)^2-m^2}\\
&\hspace{3.6cm}\times T^{\m\n\l\s}(q_1,k-q_1-q_2,q_2)\mathrm{\Pi}_{\m\n\l\s}(q_1,k-q_1-q_2,q_2),
\label{ageneral}
\end{split}
\eq
with the projector
\beq
\L_\m(p',p)=\frac{m^2}{k^2(4m^2-k^2)}\[\g_\m+\frac{k^2+2m^2}{m(k^2-4m^2)}(p'+p)_\m\],
\eq
where $m$ denotes the muon mass. 
Furthermore in Eq.~(\ref{ageneral}), the leptonic tensor $T$ is given by: 
\beq
\begin{split}
T^{\m\n\l\s}(q_1,k-q_1-q_2,q_2)&=\mathrm{Tr}\[(\notd{p}+m)\L^\s(p',p)(\notd{p}'+m)\g^\l(\notd{p}\,'-\notd{q}_2+m)\g^\n(\notd{p}+\notd{q}_1+m)\g^\m\].
\end{split}
\eq

\begin{figure}[h]
\centering
  \includegraphics[width=15cm]{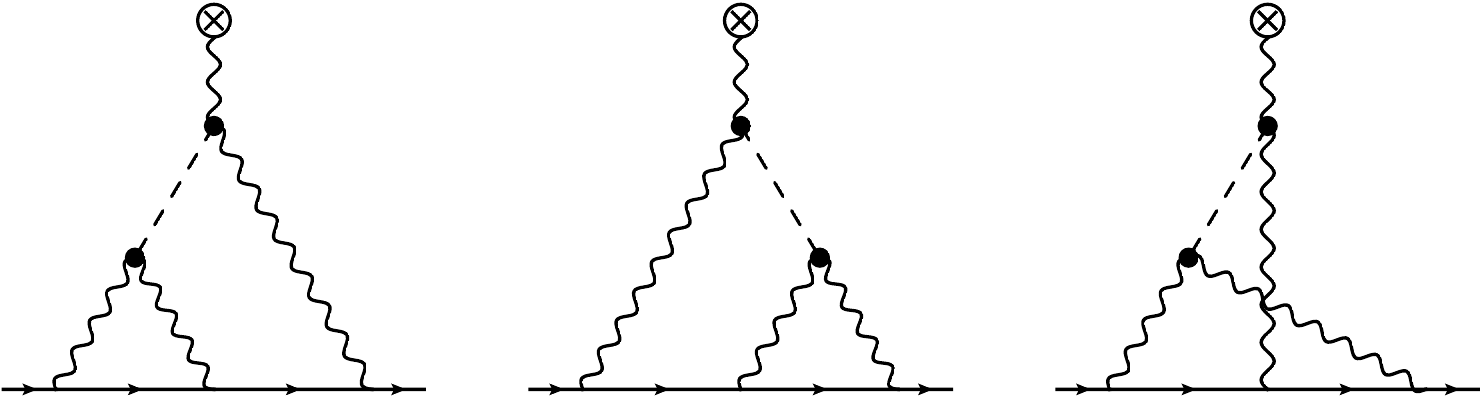}\\
  \caption{The single meson contributions to the hadronic light-by-light scattering.}
  \label{LbL_pole}
\end{figure}

The tensor $\Pi_{\m\n\l\s}(q_1,q_2,q_3)$ denotes the light-by-light vacuum polarization tensor. 
In this work, we will consider  the contributions of a single meson with an arbitrary spin to $\Pi$, 
which have the general form:
\bea
(ie)^4\Pi_{\m\n\l\s}(q_1,k-q_1-q_2,q_2) &=& {\cal M}_{\m\n,\{\a\}}(q_1,k-q_1-q_2)\frac{iP^{\{\a\},\{\b\}}(k-q_2)}{(k-q_2)^2-M^2} {\cal M}_{\l\s,\{\b\}}(q_2,-k)\nn \\
&&\hspace{-1.5cm}+ \,{\cal M}_{\m\s,\{\a\}}(q_1,-k)\frac{iP^{\{\a\},\{\b\}}(k-q_1)}{(k-q_1)^2-M^2} {\cal M}_{\n\l,\{\b\}}(k-q_1-q_2,q_2) \nn \\
&&\hspace{-1.5cm}+ \,{\cal M}_{\m\l,\{\a\}}(q_1,q_2)\frac{iP^{\{\a\},\{\b\}}(q_1+q_2)}{(q_1+q_2)^2-M^2} {\cal M}_{\n\s,\{\b\}}(k-q_1-q_2,-k), 
\label{LbL_single_meson}
\eea
where the three terms correspond with the three topologies shown in Fig.~\ref{LbL_pole}.
In Eq.~(\ref{LbL_single_meson}), 
the Lorentz amplitude $\mathcal{M}_{\m\n,\{\a\}}(q_1,q_2)$ describes the transition from the initial state of two virtual photons with momenta $q_1$ and $q_2$ to a $C$-even ($J^{P+}$) meson with mass $M$. Depending on the spin  $J$ of the meson, the amplitude $\mathcal{M}_{\m\n,\{\a\}}$ has different Lorentz structures: for the case of a pseudoscalar ($J^{PC}=0^{-+}$) and a scalar ($J^{PC}=0^{++}$) meson the amplitude is a rank two tensor, for the case of an axial-vector ($J^{PC}=1^{++}$) and a tensor ($J^{PC}=2^{++}$) meson it is a rank three tensor and a rank four tensor respectively. The projector $P$ for spin $J$ entering the meson propagator is defined by the spin sum
\beq
P^{\{\a\},\{\b\}}(p)=\sum\limits_{\s=-J}^J\ve^{\{\a\}}_\s(p)\ve_\s^{\{\b\}\ast}(p),
\label{spinprojector}
\eq
where the $\ve^{\{\a\}}_\s$ denote the corresponding polarization tensors.
In the following, we give the specific expressions which we use for the transition amplitudes and the polarization sums in our calculation. The transition amplitudes are defined in such a way that the non-perturbative physics is contained in the meson transition Form Factors (FFs). It is important to note that these FFs  depend on three invariants in the general case~\cite{Jegerlehner:2009ry, Nyffeler:2009tw}. However, mainly due to the absence of reliable information about the off-shell dependence on the virtuality of the exchanged meson we will assume, for the following estimates, the pole-dominance approximation for the FFs:
\beq
F(q_1^2,q_2^2,(q_1+q_2)^2)=F(q_1^2,q_2^2,M^2) \equiv F_{{\cal M} \g^\ast\g^\ast}(q_1^2,q_2^2), 
\eq
where $q_1^2, q_2^2$ denote the two photon virtualities, and $(q_1 + q_2)^2$ denotes the meson virtuality. 

For a \emph{pseudo-scalar meson} $({\cal P})$, the transition amplitude is defined by~:
\beq
\mathcal{M}^{({\cal P})}_{\m\n}(q_1,q_2)=-ie^2\ve_{\m\n\a\b}q_1^\a q_2^\b F_{{\cal P} \g^\ast\g^\ast}(q_1^2,q_2^2).
\label{ggPS}
\eq
The spin projection operator for $J=0$ has a trivial form~:
\beq
P(q)=1.
\label{scalar_prop}
\eq
  
A \emph{scalar meson} $({\cal S})$ may be produced either by two transverse or by two longitudinal photons~\cite{Poppe:1986dq, Schuler:1997yw}.
As the main contribution to the magnetic moment comes from the region of small photon virtualities, the contribution of the transverse amplitude is dominating. Furthermore, there is no empirical information on the structure of the longitudinal FFs at present. Thus in this work, we will only consider the transverse part of the scalar meson production amplitude which is described by:
\beq
\begin{split}
{\cal M}^{({\cal S})}_{\m\n}(q_1,q_2) &=-e^2\frac{(q_1\cdot q_2)}{M} \, R_{\m\n}(q_1,q_2) \,
F_{{\cal S} \g^\ast\g^\ast}(q_1^2,q_2^2),
\label{ggS}
\end{split}
\eq
where the symmetric transverse tensor $R^{\mu \nu}$ is defined as:
\begin{eqnarray}
R^{\mu \nu} (q_1, q_2) \equiv - g^{\mu \nu} + \frac{1}{X} \,
\bigl \{ \left(q_1 \cdot q2 \right) \left( q_1^\mu \, q_2^\nu + q_2^\mu \, q_1^\nu \right)
- q_1^2 \, q_2^\mu \, q_2^\nu  - q_2^2 \, q_1^\mu \, q_1^\nu 
\bigr \} ,
\end{eqnarray}
and $X \equiv (q_1 \cdot q_2)^2 - q_1^2 q_2^2$.
 
Although the production of an \emph{axial-vector meson} ($\cal{A}$) by two real photons is forbidden by the Landau-Yang theorem~\cite{Yang:1950rg}, 
an axial-vector meson can be produced in  two-photon processes when one or both photons are virtual. 
Existing phenomenological analyses have used an expression for the transition amplitude derived  from a non-relativistic quark model calculation~\cite{Cahn:1986qg, Achard:2001uu, Achard:2007hm}: 
\begin{align} 
{\cal M}^{({\cal A})}_{\m\n\a}(q_1,q_2) &=  ie^2 \varepsilon_{\r \nu \tau \alpha} \,  
\left[ (q_1^2g^{\r}_{\m}-q_1^\r {q_1}_\m) \, q_2^\tau -(q_2^2g^{\r}_{\m}-q_2^\r {q_2}_{\m}) \, q_1^\tau \right] \, \frac{1}{M^2} F_{{\cal A} \gamma^\ast \gamma^\ast}(q_1^2, q_2^2).
\label{eq:affcahn}
\end{align}
Note that a general discussion of the ${\cal A} \gamma^\ast \gamma^\ast$ vertex, has to allow 
for three independent Lorentz structures~\cite{Poppe:1986dq, Schuler:1997yw}. However, as no phenomenological information is available at present to disentangle the three helicity structures, 
we will use the simplified vertex of Eq.~(\ref{eq:affcahn}) in the present work. 
According to the definition of Eq.~(\ref{spinprojector}) the projection operator for spin $J=1$ is
\beq
P^{\a\b}(p)=g^{\a\b}-\frac{p^\a p^\b}{M^2}.
\eq

In this work, we also consider the two-photon production of  \emph{tensor mesons} ($\cal{T}$). 
For the light quark tensor mesons, the experimental analyses of decay angular distributions for $\g\g$ cross sections to $\p^+\p^-$, $\p^0\p^0$, $\eta\p^0$, and $K^+K^-$ channels have shown~\cite{Pennington:2008xd}  that the $J = 2$ mesons are produced predominantly (around 95\% or more) in a state of helicity $\L = 2$. We will therefore assume in all of the following analyses that the hadronic light-by-light amplitude for tensor states is dominated by the helicity $\L=2$ exchange. Therefore, the two-photon decay rate is $\Gamma_{\gamma \gamma}\left({\cal T}(\Lambda = 2) \right) \approx \Gamma_{\gamma \gamma}({\cal T})$,  and we will safely neglect the contribution of the remaining four helicity amplitudes. 
The relevant part of the Lorentz amplitude for the dominant process 
$\gamma^\ast \gamma^\ast \to {\cal T}(\Lambda=2)$,  can be parameterized as~\cite{Pascalutsa:2012pr}:
\beq
{\cal M}^{({\cal T})}_{\m\n\a\b}(q_1,q_2) 
\equiv e^2 \, \frac{(q_1\cdot q_2)}{M} \, M_{\m\n\a\b}(q_1,q_2) \, 
F_{{\cal T} \gamma^\ast \gamma^\ast}(q_1^2, q_2^2),
\label{ggT_trans}
\eq
with
\bea
&&M_{\m\n\a\b} = \left\{R_{\m\a}(q_1,q_2)R_{\n\b}(q_1,q_2)+ \frac1{8(q_1+q_2)^2\[(q_1\cdot q_2)^2-q_1^2q_2^2\]}R_{\m\n}(q_1,q_2)\right. 
\label{ggT_trans2} \\
&& 
\times \left.\[(q_1+q_2)^2(q_1-q_2)_\a-(q_1^2-q_2^2)(q_1+q_2)_\a\] 
\cdot \[(q_1+q_2)^2(q_1-q_2)_\b-(q_1^2-q_2^2)(q_1+q_2)_\b\]\right\}. 
\nonumber
\eea
The projection operator for $J=2$ has the form
\beq
P_{\a\b,\g\d}=\frac12\(K_{\a\g}K_{\b\d}+K_{\a\d}K_{\b\g}\)-\frac13K_{\a\b}K_{\g\d},
\label{tensor_prop}
\eq
with
$K_{\a\b} \equiv -g_{a\b}+ p_\a p_\b / p^2 $.

By an appropriate change of variables one can show that the first and second diagrams of Fig.~\ref{LbL_pole} give equal contributions to Eq.~(\ref{ageneral}). Thus, the two-loop integral for 
$a_\mu$ may be defined as a sum of two terms~:
\beq
\begin{split}
a_\m^{LbL}=\lim\limits_{k\to0}-e^6&\int\frac{\dd^4 q_1}{(2\p)^4}\int\frac{\dd^4 q_2}{(2\p)^4}\frac{1}{q_1^2q_2^2(k-q_1-q_2)^2\[(p+q_1)^2-m^2\]\[(p+k-q_2)^2-m^2\]}\\
\times&\[\frac{F_{{\cal M}\g^\ast\g^\ast}(q_1^2,(k-q_1-q_2)^2)F_{{\cal M} \g^\ast\g^\ast}(k^2,q_2^2)}{(k-q_2)^2-M^2}T_{1}(q_1,k-q_1-q_2,q_2)\right.\\
&+\left.\frac{F_{{\cal M} \g^\ast\g^\ast}(q_1^2,q_2^2)F_{{\cal M} \g^\ast\g^\ast}((k-q_1-q_2)^2,k^2)}{(q_1+q_2)^2-M^2}T_{2}(q_1,k-q_1-q_2,q_2)\].
\end{split}
\label{amm}
\eq
Here, $T_1$ is given by the contraction of the first two terms in Eq.~(\ref{LbL_single_meson}) 
with the tensor $T^{\m\n\l\s}$; whereas $T_2$ is defined by the contraction of $T^{\m\n\l\s}$ 
with the third term.
Computing the Dirac traces (for which we used the computer algebraic system FORM~\cite{Kuipers:2012rf}), 
we find that $T_1$ and $T_2$ contain a set of structures of three types~:
\begin{align}
(q_1\cdot k)^i(q_2\cdot k)^j,\quad\frac{(q_1\cdot k)^i(q_2\cdot k)^j}{(q_2\cdot k)^2-q_2^2k^2},\quad\frac{(q_1\cdot k)^i(q_2\cdot k)^j}{\[(q_2\cdot k)^2-q_2^2k^2\]^2}.\nn
\end{align}

Before taking the limit $k\to0$, we first need to eliminate the dependence on the spatial direction of $k$. Since the trace under consideration projects to a scalar, we may average the $\Omega(\hat{\mathbf{k}})$ dependence over all spatial directions without changing the result for $a_\m$:
\beq
\int\frac{\dd\Omega (\hat{\bold{k}})}{4\p}\;a_\m(\bold{k})=a_\m .
\label{averaging}
\eq
After taking the limit $k \to 0$ explicitely, 
we integrate the angular dependence on $\Omega(\hat{\mathbf{q}}_1)$ and $\Omega(\hat{\mathbf{q}}_2)$. 
For the angular integrations, we use the technique based on the properties of the Legendre polynomials. Given a particular parametrization of the FFs, the angular integrals may be performed analytically. We will give more technical details in a forthcoming publication. In this work, we will use both monopole (mon) and dipole (dip) parameterizations of the form~:
\bea
\frac{F_{\cal{M} \gamma^\ast \gamma^\ast}^{\mathrm{mon}} \(q_1^2, q_2^2\)}{F_{\cal{M} \gamma^\ast \gamma^\ast} \(0,0\)} &=&\frac1{\(1- q_1^2/\Lambda_{\mathrm{mon}}^2\)}\frac1{\(1- q_2^2/\Lambda_{\mathrm{mon}}^2\)} \, , 
\label{eq:monopole} \\
\frac{F_{\cal{M} \gamma^\ast \gamma^\ast}^{\mathrm{dip}} \(q_1^2, q_2^2\)}{F_{\cal{M} \gamma^\ast \gamma^\ast} \(0,0\)} &=&\frac1{\(1- q_1^2/\L_{\mathrm{dip}}^2\)^2} \, \frac1{\(1- q_2^2/\L_{\mathrm{dip}}^2\)^2} \, ,
\label{eq:dipole}
\eea
where $\Lambda_{\mathrm{mon}}$ ($\Lambda_{\mathrm{dip}}$)
are the monopole (dipole) mass parameters respectively, 
which are to be determined from phenomenology. 

After working out the angular integrations analytically, we perform the remaining integrals numerically. For convenience, we perform a Wick rotation for the energy component of the four-momenta $q_1$ and $q_2$ and carry out the numerical integration in polar coordinates. In particular, we make the change of variables~:
\beq
Q^0_i=Q_i\cos\psi_i,\qquad |\mathbf{Q}_i|=Q_i\sin\psi_i, 
\eq
where $Q_i^2 \equiv - q_i^2$.
As an example, the resulting four-dimensional integral for the case of a monopole FF takes the form~:
\bea
a_\m^{LbL} &=&-\frac{4\a^3}{\p^3}(2J+1)|F_{{\cal M}\g^\ast\g^\ast}(0,0)|^2\int\limits_0^\infty\dd Q_1\int\limits_0^{\p}\dd \psi_1\,\,\int\limits_0^\infty\dd Q_2\int\limits_0^{\p}\dd \psi_2\frac1{Q_1^2/\L^2+1}\frac1{Q_2^2/\L^2+1} \nonumber \\
&\times& \frac1{Q_2^2+M^2}\frac{\sin^2\psi_1\sin^2\psi_2}{Q_1+2im\cos\psi_1}\[2\;\frac{\tilde{T}_{1}(Q_1,Q_2,\psi_1,\psi_2)}{Q_2-2im\cos\psi_2}+Q_2\tilde{T}_{2}(Q_1,Q_2,\psi_1,\psi_2)\],
\label{amuscalar4dim}
\eea
where the factor $\[(Q_1+Q_2)^2 +\L^2\]^{-1}$ is absorbed in the expressions for $\tilde{T}_1$ and 
$\tilde{T}_2$ and in the third diagram we have absorbed a factor 
$\[(Q_1+Q_2)^2+2im(Q_1 \cos\psi_1+ Q_2 \cos {\psi_2})\]^{-1}$ into $\tilde{T}_2$.

\section{Result and discussion}

To test our formalism, we have firstly applied it to the case of pseudo-scalar meson poles. 
This case had been worked out in Ref.~\cite{Knecht:2001qf} using the Gegenbauer polynomial technique, 
where for pole parametrizations of the FFs the HLbL contribution to $a_\mu$ had been given by a two-dimensional numerical integral over $Q_1$ and $Q_2$. We checked that using e.g. a monopole FF, the result obtained from Eq.~(\ref{amuscalar4dim}) is in exact agreement with the result of Ref.~\cite{Knecht:2001qf}.   Due to the more complicated vertex structure for axial-vector, scalar and tensor mesons, the Gegenbauer polynomial technique cannot be easily extended, which is why we resort to the four-dimensional expression of Eq.~(\ref{amuscalar4dim}).  
Using this formalism, we subsequently discuss our estimates for the HLbL contribution to $a_\mu$ due to axial-vector, scalar and tensor mesons.  

\subsection{Axial-vector mesons}

For an axial-vector meson, it is conventional to define an equivalent two-photon 
decay width to describe its decay into one quasi-real longitudinal photon (with virtuality $Q_1^2$) and a transverse (real) photon as~\cite{Schuler:1997yw, Achard:2001uu, Achard:2007hm}:
\begin{eqnarray}
\tilde \Gamma_{ \gamma \gamma} \equiv \lim \limits_{Q_1^2 \to 0} \, \frac{M^2}{Q_1^2} \, \frac{1}{2} \,
\Gamma \left( {\cal{A}} \to \gamma^\ast_L \gamma_T \right), 
\label{a2gwidth}
\end{eqnarray}
which allows to express the FF normalization entering 
the $\cal{A} \gamma^\ast \gamma^\ast$ vertex of Eq.~(\ref{eq:affcahn}) as~\cite{Pascalutsa:2012pr}:
\beq
\[F_{\cal{A} \gamma^\ast \gamma^\ast}\(0,0\)\]^2=\frac{3}{M} \frac{4}{\p\a^2}\tilde{\G}_{\g\g}.  
\eq

Phenomenologically, the two-photon production cross sections have been measured for the two lowest lying axial-vector mesons~: 
$f_1(1285)$ and $f_1(1420)$. The most recent measurements were performed by the L3 Collaboration~\cite{Achard:2001uu, Achard:2007hm}. 
In those works, the non-relativistic quark model expression of Eq.~(\ref{eq:affcahn}) in terms of a single FF $F_{\cal{A} \gamma^\ast \gamma^\ast}$ has been assumed, 
and the resulting FF has been parameterized by a dipole form as 
in Eq.~(\ref{eq:dipole}). 
Table~\ref{tab_ax} shows the present experimental status of the equivalent 
$2 \gamma$ decay widths for $f_1(1285)$, and $f_1(1420)$, as well as the phenomenological values for 
the dipole mass parameters $\Lambda_{\mathrm{dip}}$. 

Using these values, we can calculate the HLbL contributions of 
$f_1(1285)$ and $f_1(1420)$ to $a_\mu$, which are shown in 
Table~\ref{tab_ax}. Both contributions sum up to a value of 
$6.4 \times 10^{-11}$, which is roughly one order of 
magnitude smaller than the dominant HLbL contribution to $a_\mu$ due to the $\pi^0$~\cite{Knecht:2001qf}. 
We like to emphasize that our estimate for the two dominant axial-vector meson contributions is based on available experimental information.  In this way, we are also able to provide an error estimate, which derives from the experimental uncertainties in the equivalent $2 \gamma$ decay widths and from the FF parameterization. 

\begin{table}[h]
{\centering \begin{tabular}{|c|c|c|c|c|c|c|}
\hline
& $M$   & $\tilde \Gamma_{\gamma \gamma} $  & $\Lambda_{\mathrm{dip}}$  &   $ a_\mu $  \\
&  [MeV] &  [keV]  & [MeV]  &  [$10^{-11}$]  \\
\hline 
\quad $f_1 (1285)$  \quad & \quad $1281.8 \pm 0.6$ \quad   & \quad  $3.5 \pm 0.8 $  \quad  & \quad  $1040 \pm 78 $  \quad  
&  \quad $ 5.0  \pm 2.0 $  \quad  \\
\quad $f_1 (1420)$ \quad  & \quad $1426.4 \pm 0.9$  \quad  & \quad  $ 3.2 \pm 0.9 $ \quad  & \quad  $ 926 \pm 78 $ \quad 
&  \quad $ 1.4  \pm 0.7 $  \quad  \\
\hline
\hline
Sum  &  & & 
&  \quad $ 6.4  \pm 2.0 $  \quad  \\
\hline 
\end{tabular}\par}
\caption{Present values~\cite{PDG} of the 
$f_1(1285)$  meson~ and $f_1(1420)$ meson masses $M$, their 
equivalent $2 \gamma$ decay widths $\tilde \Gamma_{\gamma \gamma}$, defined according to Eq.~(\ref{a2gwidth}), as well as their 
dipole masses $\Lambda_{\mathrm{dip}}$ 
entering the FF of Eq.~(\ref{eq:affcahn}). 
For $\tilde \Gamma_{\gamma \gamma}$, we use the experimental results from the L3 Collaboration~:  
$f_1(1285)$ from~\cite{Achard:2001uu}, 
$f_1(1420)$ from~\cite{Achard:2007hm}. }
\label{tab_ax}
\end{table}

\begin{figure}[h]
  \includegraphics[width=8.5cm]{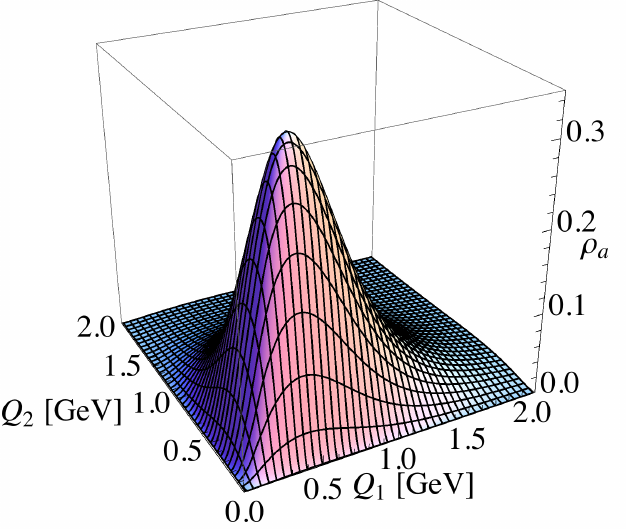}
  \includegraphics[width=8.5cm]{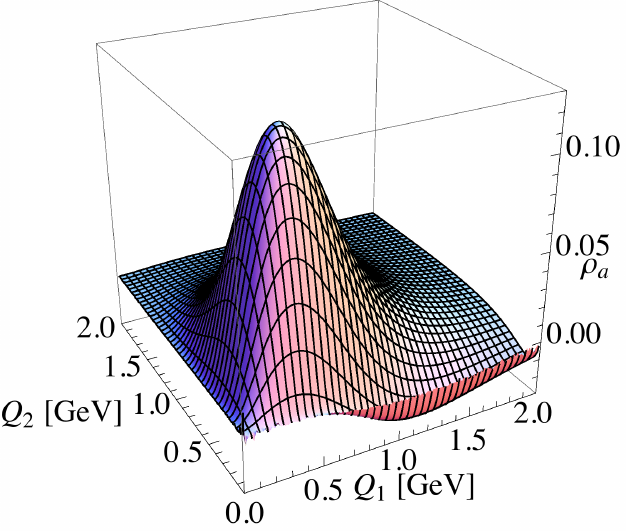}
  \caption{The density $\rho_a$ as defined in Eq.~(\ref{eq:densamu}), in units $10^{-10}$ GeV$^{-2}$, for the axial-vector meson $f_1(1285)$. Left panel corresponds with left two diagrams of Fig.~\ref{LbL_pole}, 
  right panel corresponds with right diagram of Fig.~\ref{LbL_pole}.}
  \label{axial_density}
\end{figure}

In order to have a better understanding which region of virtualities in the axial-vector meson FFs is contributing mostly to this result, it is instructive 
to define a density function $\rho_a$ as~:
\begin{equation}
a_\mu^{LbL} = \int_0^{\infty}  dQ_1 \int_0^{\infty} dQ_2 \, \rho_a (Q_1, Q_2).
\label{eq:densamu} 
\end{equation}
We show the dependence of $\rho_a$ on the photon virtualities $Q_1$ and $Q_2$, which enter the HLbL scattering diagram, for the axial-vector meson $f_1(1285)$ in Fig.~\ref{axial_density}. One notices that the dominant contribution arises 
from the region around $Q_1 \approx Q_2 \approx 0.5$ GeV. 
One also sees that the contribution beyond $Q_{1, 2} \geq 1.5$ GeV becomes negligible.  

\subsection{Scalar mesons}

We next turn to the estimate for scalar mesons ($\cal{S}$). 
The normalization of the FF corresponding with two transverse photons is related to the 
two-photon decay width of the scalar meson as~\cite{Pascalutsa:2012pr}:
\beq
\[F_{\cal{S} \gamma^\ast \gamma^\ast}\(0,0\)\]^2=\frac{1}{M} \frac{4}{\p\a^2}\, {\G}_{\g\g}.  
\eq
When going to virtual photons, unfortunately no empirical information is 
available at present for the $\cal{S} \gamma^\ast \gamma^\ast$ 
transition FFs. We will assume a simple monopole behavior of the FF. The monopole mass $\L_{\mathrm{mon}}$ is considered as a free parameter, which we will vary in the expected hadronic range $\L_{\mathrm{mon}} = 1 - 2$~GeV, 
in order to obtain the numerical estimates for $a_\mu$. 
We show our results for the HLbL contribution to $a_\mu$ due to 
the leading scalar mesons $f_0, f_0^\prime$, and $a_0$ in 
Table~\ref{table_scalar}. We find a negative contribution of the scalar 
mesons to $a_\mu$ which is in the range $-1$ to $-3$ (in units $10^{-11}$),  
when varying $\L_{\mathrm{mon}}$ in the range $1$ to $2$ GeV.

Note that in this work we do not include the contribution from the low-lying and very broad $f_0(500)$ state,  
which requires a full treatment of the HLbL contribution to $a_\mu$ due to $2 \pi$ intermediate states. 
Such treatment goes beyond the pole model framework described here.

\begin{table}[h]
{\centering \begin{tabular}{|c|c|c|c|c|}
\hline
& $M$  & $\Gamma_{\gamma \gamma} $  &   $ a_\mu $ ($\Lambda_{\mathrm{mon}}$ = 1 GeV) &   $ a_\mu $ ($\Lambda_{\mathrm{mon}}$ = 2 GeV) \\
&  [MeV] &  [keV] &  [$10^{-11}$]  & [$10^{-11}$]  \\
\hline 
\hline
\quad $f_0(980)$ \quad  & \quad $980 \pm 10$ \quad   & \quad  $0.29 \pm 0.07$  \quad   &  \quad $ -0.19 \pm 0.05 $  \quad &  \quad $ -0.61 \pm 0.15 $ \quad \\
\quad $f_0^\prime(1370)$  \quad & \quad $1200 - 1500 $  \quad  & \quad  $3.8 \pm 1.5$ \quad &  \quad $ -0.54 \pm 0.21 $ \quad &  \quad $ -1.84 \pm 0.73 $ \quad  \\
\quad $a_0(980)$  \quad &  \quad $980 \pm 20$  \quad  & \quad  $0.3 \pm 0.1$ \quad  &  \quad $ -0.20 \pm 0.07 $ \quad &  \quad $ -0.63 \pm 0.21 $ \quad  \\
\hline
\hline
Sum &  & &  $-0.9 \pm 0.2 $ & $-3.1 \pm 0.8$ \quad \\
\hline
\end{tabular}\par}
\caption{Scalar meson pole contribution to $a_\mu$ based on the present PDG values~\cite{PDG} of the scalar meson masses $M$ and their $2 \gamma$ decay widths $\Gamma_{\gamma \gamma}$.  }
\label{table_scalar}
\end{table}

\subsection{Tensor mesons}

In this work, we also estimate the HLbL contribution to $a_\mu$ due to tensor mesons ($\cal{T}$). 
The dominant tensor mesons produced in two-photon fusion processes are given by~: 
$f_2(1270)$, $a_2(1320)$, $f_2(1565)$, and $a_2(1700)$, see 
Table~\ref{table_tensor}.  
As described above, we will assume in our analysis that the tensor meson is only produced in a state of helicity 2. This allows to express the normalization of the dominant (helicity-2) FF entering the $\cal{T} \gamma^\ast \gamma^\ast$ vertex as~\cite{Pascalutsa:2012pr}:
\beq
\[F_{\cal{T} \gamma^\ast \gamma^\ast}\(0,0\)\]^2=\frac{5}{M} \frac{4}{\p\a^2}\, {\G}_{\g\g}.  
\eq
At the present moment there is unfortunately no direct experimental information about the $Q^2$ dependence of the tensor meson transition FFs. One can however resort to other phenomenological information based on exact forward sum rules for the light-by-light scattering. 
For $\gamma^\ast \gamma \to X$ fusion cross sections, with one real photon ($\gamma$) and one virtual photon ($\gamma^\ast$), three super convergence sum rules were derived in Refs.~\cite{Pascalutsa:2010sj, Pascalutsa:2012pr}. 
Applied to the $\gamma^\ast \gamma$ production of mesons, this leads to intricate relations between transition FFs of pseudo-scalar, 
axial and tensor mesons. In order to saturate these sum rules, 
one obtains approximate expressions for the dominant tensor meson transition FFs, given the knowledge of the transition FFs for the pseudo-scalar mesons. In particular, it was shown in Ref.~\cite{Pascalutsa:2012pr} that the $\eta$, and $\eta^\prime$ transition FFs constrain the transition FF for  $f_2(1270)$ and the $\pi^0$ transition FF constrains the corresponding transition FF for the $a_2(1320)$ state. 
We found that these relations can approximately 
be expressed by choosing a dipole form for the tensor meson transition FF with dipole mass parameter $\Lambda_{\mathrm{dip}} = 1.5$~GeV. 
We use this estimate in calculating the HLbL contribution to 
$a_\mu$ due to tensor mesons, which is shown in 
Table~\ref{table_tensor}. We see that the four dominant 
tensor meson contributions add up to a contribution to $a_\mu$ of around $1$ (in units $10^{-11}$).

\begin{table}[h]
{\centering \begin{tabular}{|c|c|c|c|}
\hline
& $M$  & $\Gamma_{\gamma \gamma} $  &  $ a_\mu $ ($\Lambda_{\mathrm{dip}}$ = 1.5 GeV)    \\
&  [MeV] &  [keV] &  [$10^{-11}$]  \\
\hline 
\hline
\quad $f_2 (1270)$ \quad & \quad $1275.1 \pm 1.2 $ \quad   & \quad  $3.03 \pm 0.35$  \quad  & \quad $ 0.79 \pm 0.09$ \quad   \\
\quad $f_2 (1565)$ \quad & \quad $1562 \pm 13 $ \quad   & \quad  $0.70 \pm 0.14$  \quad &  
\quad $ 0.07 \pm 0.01$ \quad  \\
\hline
\quad $a_2 (1320)$ \quad &  \quad $1318.3 \pm 0.6$  \quad  & \quad  $1.00 \pm 0.06$ \quad   & \quad  $ 0.22 \pm 0.01 $  \quad  \\
\quad $a_2 (1700)$ \quad &  \quad $1732 \pm 16$  \quad  & \quad  $0.30 \pm 0.05$ \quad  &  
\quad $ 0.02 \pm $ 0.003 \quad  \\
\hline
\hline
Sum &  & &  \quad $ 1.1  \pm 0.1$ \quad \\
\hline
\end{tabular}\par}
\caption{Tensor meson pole contribution to $a_\mu$ based on the present PDG values~\cite{PDG} of the tensor meson masses $M$ and their $2 \gamma$ decay widths $\Gamma_{\gamma \gamma}$.  
}
\label{table_tensor}
\end{table}

\subsection{Comparison with previous works}

Our results can be compared with previous estimates for 
axial-vector and scalar mesons, which are shown in Table~\ref{tab_comp}. For tensor mesons, our results are the first estimates. 

The previous estimates for axial-vector mesons differ quite a lot. 
The BPP estimate~\cite{BPP} is based on an extended Nambu-Jona-Lasinio model in which both a $1/N_c$ and chiral counting was used. 
The HKS estimate~\cite{HKS, HK} for axial-vector meson FFs was 
based on a hidden local gauge symmetry model. 
The MV estimate~\cite{MV}, which was also adopted in JN~\cite{Jegerlehner:2009ry} is an order of magnitude larger than the 
BPP and HKS estimates, and around a factor 3 larger than our estimate. 
The large value obtained in Ref.~\cite{MV} was obtained because a constant FF was used at the external vertex to reproduce the QCD short-distance constraints. Although such short-distance constraints are surely important for the large $Q^2$ behavior of the 
FFs, one can see from Fig.~\ref{axial_density} that using the 
empirical information for the $f_1(1285)$ transition FF, the 
region which dominates the HLbL contribution to $a_\mu$ is 
for virtualities around and below 1 GeV$^2$. It has furthermore been argued by PdRV \cite{Prades:2009tw} that the errors in the BPP and HKS estimates were underestimated, and an intermediate estimate with larger error has been suggested, which is in agreement within $1 \sigma$ with our estimate. 

For the scalar mesons, BPP has performed an estimate,  
which was adopted by N/JN and PdRV (by increasing the error bar to 
100 \%). Compared with the result of BPP, our result also has the negative sign, but is around a factor of 2 smaller in magnitude. 
Given that there is no empirical information at all on the scalar meson transition FFs, future data from BES-III would be mostly welcome here to better constrain this contribution. 
  
\begin{table}[h]
{\centering \begin{tabular}{|c|c|c|c|c|c|}
\hline
& axial-vectors   &  scalars  & tensors   \\
\hline 
BPP  \cite{BPP}  & \quad $2.5 \pm 1.0$ \quad   & \quad $-7 \pm 2$ \quad & \quad  -  \quad    \\
HKS \cite{HKS, HK}  & \quad $1.7 \pm 1.7$ \quad   & \quad - \quad & \quad  -  \quad    \\
MV \cite{MV} & \quad $22 \pm 5$ \quad   & \quad - \quad & \quad  -  \quad    \\
PdRV \cite{Prades:2009tw}  & \quad $15 \pm 10$ \quad   & \quad $-7 \pm 7$ \quad & \quad  -  \quad    \\
N/JN \cite{Jegerlehner:2009ry} & \quad $22 \pm 5$ \quad   & \quad $-7 \pm 2$ \quad & \quad  -  \quad    \\
\hline
\hline
this work  & \quad $6.4 \pm 2.0$ \quad   & \quad $-3.1 \pm 0.8$ \quad & \quad  $1.1 \pm 0.1 $  \quad    \\
\hline 
\end{tabular}\par}
\caption{HLbL contribution to $a_\mu$ (in units $10^{-11}$)
due to axial-vector, scalar, and tensor mesons obtained in this work,
compared with various previous estimates. 
For our scalar meson estimate, we have quoted the value corresponding with $\Lambda_{\mathrm{mon}} = 2$ GeV. }
\label{tab_comp}
\end{table}

\section{Conclusions}

In this letter we have presented the formalism to calculate the HLbL contribution to the muon's anomalous 
magnetic moment $a_\mu$ due to axial-vector, scalar and tensor meson poles. In this way, we have extended the framework 
of Ref.~\cite{Knecht:2001qf}, where the leading HLbL contribution due to pseudo-scalar mesons 
was evaluated. To allow for the different Lorentz structures of the $\gamma^\ast \gamma^\ast \to$ meson vertex, we have 
performed a combined analytical and numerical technique, where the angular integrals over the virtual photon 
momenta were performed analytically using the Legendre polynomial technique, and where the resulting 
four-dimensional integral for $a_\mu$ was performed numerical. We validated our method by reproducing the 
known result for pseudo-scalar mesons. To estimate the HLbL contribution to $a_\mu$ from axial-vector, scalar and tensor mesons, we incorporated available experimental input as well as constraints from 
light-by-light scattering sum rules. For those mesons which have the largest known couplings to two virtual photons, we 
obtained as estimates~: 
\begin{eqnarray}
a_\mu (f_1, f_1^\prime) &=& (6.4 \pm 2.0) \times 10^{-11}, 
\nonumber \\ 
a_\mu (f_0, f_0^\prime, a_0) &=& [(-0.9 \pm 0.2) \; \mathrm{to} \; (-3.1 \pm 0.8)] \times 10^{-11}, 
\nonumber \\ 
a_\mu (f_2, f_2^\prime, a_2, a_2^\prime) &=& (1.1 \pm 0.1) \times 10^{-11}.
\nonumber
\end{eqnarray}
The size of such contributions is about an order of magnitude smaller than the dominant $\pi^0$ HLbL contribution. 
Given a new muon $g-2$ experiment at Fermilab, which aims to reduce the experimental uncertainty 
to $\delta a_\mu \approx 16 \times 10^{-11}$, it is however crucial to further constrain the 
theoretical uncertainty due to the HLbL contribution. 
In this respect, it would be particularly helpful to have  $\gamma^\ast \gamma^\ast \to$ meson transition form factor measurements with one and two virtual photons for axial-vector, scalar and tensor states. 
As such information will become available, in particular from future measurements from BES-III, the here developed formalism can be used to further improve on the estimate of the HLbL contribution to $a_\mu$.

\section*{Acknowledgements}

We like to thank A. Denig and P. Masjuan for useful discussions. 
This work was supported by the Deutsche Forschungsgemeinschaft DFG in part through the Collaborative 
Research Center ``The Low-Energy Frontier of the Standard Model" (SFB 1044), 
in part through the graduate school Graduate School ``Symmetry Breaking in Fundamental Interactions" 
(DFG/GRK 1581), and in part through 
the Cluster of Excellence "Precision Physics, Fundamental Interactions and Structure of Matter" (PRISMA).

\end{document}